\documentclass[12pt]{article}
\usepackage{setspace}
\usepackage{amsmath}
\usepackage[top=1in,bottom=1in,right=1in,left=0.5in]{geometry}
\usepackage{caption}
\usepackage{psfrag}
\usepackage{graphicx}
\usepackage{hyperref}
\usepackage{subfigure}
\usepackage{authblk}
\usepackage{verbatim}
\usepackage{multicol}
\usepackage{bibcheck}
\usepackage{comment}
\usepackage{float}
\usepackage[natbib=true,sorting=none,style=nature]{biblatex}
\title{Transport properties of baryon rich back-reacted thermal plasma with finite 't Hooft coupling correction}
\author{Rishi Pokhrel\thanks{E-Mail: rishipokhrel.smit@gmail.com and rishi\_20211037@smit.smu.edu.in} }
\author{Indra K. P. Chettri\thanks{E-Mail: indrapandey.smit@gmail.com and indra\_202410054@smit.smu.edu.in} }
\author{Karma P. Sherpa\thanks{E-Mail: sherpa.karma.pincho@gmail.com and karma\_202310028@smit.smu.edu.in} }
\author{Tanay K. Dey\thanks{E-mail: tanay.dey@gmail.com and tanay.d@smit.smu.edu.in}}
\affil{Department of Physics, Sikkim Manipal Institute of Technology, Sikkim Manipal University, Majitar, Rangpo, East Sikkim, 737136, India.}

\date{}
\begin{document}
	\maketitle

	\begin{abstract}
		In this work, holographic approach has been used to analyse the transport properties of baryon rich back-reacted thermal plasma with finite 't Hooft coupling correction. The dual bulk geometry is charged AdS black hole with higher derivative Gauss-Bonnet (GB) correction and string cloud. Specially, we have studied the nature of drag force, jet quenching parameter, screening length, radial profile and energy loss with respect to different parameters. The drag force and jet quenching parameter are enhanced with GB coupling, baryon and flavor density whereas the screening length reduces with these parameters. The radial profile and energy loss of the rotating quark has also been studied and it is observed that the radial profile decreases with increase in baryon potential and flavor density, temperature and angular frequency, whereas it is enhanced with conjugate momenta and GB coupling. Further, the energy loss of the quark grows with potential and flavor density, velocity and angular frequency and it is suppressed with GB coupling. 
	\end{abstract}

\clearpage


\section{Introduction}

The gauge/gravity duality provides a powerful nonperturbative framework to study strongly coupled gauge theories via classical gravitational dynamics in asymptotically Anti-de Sitter (AdS) spacetimes \cite{Maldacena1998,Maldacena1999,Witten1998}. Since its introduction, the correspondence has found wide-ranging applications in exploring thermal and transport properties of strongly interacting plasmas, particularly in contexts where conventional field-theoretic techniques are inadequate. One of its most prominent applications lies in modelling aspect of the Quark-Gluon Plasma (QGP) produced in ultra-relativistic heavy-ion collisions at RHIC and the LHC, where experimental evidence points toward the formation of a deconfined medium of quarks and gluons exhibiting strong collective behaviour \cite{Zajc_2008, muller2007,Shuryak_2007, d_Enterria_2007, salgado2006, Shuryak_2007_report,Tannenbaum_2006, Muller_2006,Gyulassy_2005, Adcox_2005, Back_2005, Adams_2005,Baier_1997, Eskola_2005}.
A striking outcome of heavy-ion phenomenology is the near-perfect fluid nature of the QGP, characterized by a very small shear viscosity to entropy density ratio \cite{Sadeghi:2022kgi}. Hydrodynamical studies incorporating low viscosity have been successful in reproducing experimental observables, motivating extensive holographic investigations of transport coefficients in strongly coupled plasmas \cite{Kovtun:2004de,Son:2007vk,Romatschke:2007mq,Song:2007fn,Dusling:2007gi,PHENIX:2006iih,Romatschke:2007eb}.

In the holographic description, a heavy external probe quark propagating through a thermal plasma is represented by the endpoint of an open string stretching from the AdS boundary toward the black hole horizon and the hanging body of the string into the AdS bulk representing the gluonic field between the quark-antiquark pair ($q\bar{q}$). The resulting trailing string configuration captures the dissipation of energy and momentum into the medium and leads to a finite drag force acting on the quark \cite{Herzog_2006,Gubser_2006,Caceres2006,Casalderrey_Solana_2006,Matsuo_2006,Talavera_2007,Nakano_2007,Chen_2024}. Closely related observables include the jet-quenching parameter ($\hat{q}$), which characterizes transverse momentum broadening of energetic partons and can be computed holographically using light-like Wilson loops \cite{Liu:2006ug,Caceres2006a,Nakano_2007,Bertoldi_2007,Cotrone_2007,Bigazzi_2009,Bigazzi_2011,Chen_2024}. In addition, the screening length of a moving $q\bar{q}$ pair provides insight into colour screening and quarkonium dissociation in the plasma \cite{Chernicoff_2006,Chernicoff2008,Liu_2007a,Zhu_2024}. Further, considering a rotating probe, the radial profile and energy loss has been studied in \cite{Fadafan_2009,Athanasiou_2010,Herzog_2006,Herzog_2007,Mikhailov2003,Hou_2021,Fadafan_2008,Atashi_2020}.
While early holographic studies focused primarily on plasmas in the infinite coupling limit, considerable effort has been made to include geometries that makes the setup more QCD like. One such ingredient is the inclusion of fundamental flavor degrees of freedom at finite density. When the number of flavors is large, their back reaction on the geometry becomes significant and can be effectively modelled by a uniformly distributed cloud of fundamental strings extending from the boundary to the horizon. The Presence of string cloud leads to a modified black hole geometry that captures essential features of a flavor-rich thermal plasma \cite{Chakrabortty2011a,Chakrabortty2016a}.
Another crucial aspect relevant to holographic studies is the baryon potential. Holographically, finite baryon potential is introduced through charged AdS black hole \cite{Pokhrel2023}. Recent studies have combined the effects of a string cloud with a charged geometry, providing a holographic dual of a back-reacted baryon rich thermal plasma and exploring its thermodynamical properties \cite{Pokhrel_2025}. These analyses reveal that baryon potential and flavor back-reaction can significantly alter transport behaviour and probe dynamics.
Beyond these structural deformations, finite 't Hooft coupling effects are encoded holographically through higher-derivative terms in the bulk action, with Gauss-Bonnet (GB) gravity providing a particularly tractable framework \cite{Dey:2020yzl}. Finite-coupling corrections have been shown to modify drag force, energy loss mechanisms and screening properties of heavy probes \cite{Fadafan_2008,Dey:2020yzl,Atashi_2020,Pokhrel_2025a}. They also impact the shear viscosity bound, leading to deviations from the universal KSS value within allowed parameter ranges \cite{Kovtun:2004de,Son:2007vk} which supports the recent data of RHIC \cite{Romatschke:2007mq,Song:2007fn,PHENIX:2006iih,Romatschke:2007eb,Dusling:2007gi} suggesting a lower value of viscosity to entropy density ratio ($\frac{\eta}{s}$). Which indicates the strongly coupled QGP produced in the heavy ion collision experiments can be treated as a perfect fluid having small viscosity to entropy density ratio.

In this work, we investigate the transport properties of a back-reacted baryon rich thermal plasma with finite 't Hooft coupling corrections. Our holographic setup simultaneously incorporates a finite baryon potential, flavor back-reaction and 't Hooft coupling via charge, string cloud and higher-derivative GB corrections respectively \cite{Dey:2023inw}. Within this unified framework, we analyze the drag force and energy loss of a moving heavy quark, the jet-quenching parameter and the screening length of a moving $q\bar{q}$ pair, as well as rotating probe configurations. Our results elucidate how the combined effects of baryon potential, flavor back-reaction and finite coupling corrections affects the transport properties of strongly coupled thermal plasma, offering a more QCD-motivated holographic perspective.

The drag force has been analyzed, indicating an enhancement in parameters lead to an increase in the experienced drag, whereas for high temperature and velocity region the drag force reduces slightly with increase in GB coupling. The jet quenching parameter has been computed which shows that increase in temperature, flavor density, GB coefficient and baryon potential leads to the enhancement of the jet quenching parameter. This indicates that the energy loss of the probe quark due to the suppression of the high transverse momentum in the thermal medium increases with these parameters. Investigation of the screening length of the $q\bar{q}$ pair in both perpendicular and parallel orientation showed that increment in rapidity parameter, temperature, flavor density, potential and finite 't Hooft coupling led to a decrease in the screening length. A comparison of the values of screening length in both orientations showed that the screening length in parallel orientation is greater than that of perpendicular orientation for the same set of parameters demanding a more stable bound state configuration in parallel orientation. The study of the radial profile of the rotating string showed that the radial profile is inversely related to baryon potential and flavor density, temperature and angular frequency whereas it is enhanced with increasing conjugate momenta and finite 't Hooft coupling. Further, the rotational energy loss is examined revealing that it grows with velocity, angular frequency, baryon and flavor density, however it is suppressed with finite 't Hooft coupling. Similarly, the drag energy loss is studied which shows consistent qualitative behaviour with that of rotational energy loss.

The work is organised as follows: In section (\ref{sec_dual_bulk}), we discuss the dual bulk geometry of the baryon rich back-reacted thermal plasma with finite 't Hooft coupling correction. In section (\ref{sec_drag_force}), we compute the drag force experienced by an external probe quark translating through the baryon rich back-reacted thermal plasma with finite 't Hooft coupling correction and the jet quenching parameter is studied in sec (\ref{sec_jetQ}). In section (\ref{sec_screening_length}), we study the screening length of the quark-antiquark pair in both perpendicular and parallel orientation. In section (\ref{sec_radial_profile}) and section (\ref{sec_energy_loss}), we compute the radial profile and energy loss of the rotating quark respectively. Finally, we summarise our findings in section (\ref{sec_conclusion}).

\section{Dual Bulk}
\label{sec_dual_bulk}
In this section, we briefly discuss the dual bulk geometry of baryon rich back-reacted thermal plasma with finite 't Hooft coupling described by the charged AdS black hole solution with higher derivative GB correction and string cloud density respectively. The bulk gravity action is given by,
\begin{equation}
	 S = -\frac{1}{16\pi G_5}\int d^{4+1}x \sqrt{-g} \left[R-2\Lambda +\alpha \hat{L} -F^2\right] + S^{SC},
	 \label{eq_bulk_action}
\end{equation}
where $G_5$, $g_{\mu\nu}$, $R$, $\Lambda$, $\alpha$ and $F=F_{\mu\nu}F^{\mu\nu}$ are the five dimensional Gravitational constant, bulk metric tensor, Ricci scalar, cosmological constant, GB coefficient and Maxwell field strength respectively. The GB term $\hat{L}$ is given by,
\begin{equation}
    \hat{L} = R^2 - 4 R_{\mu\nu}R^{\mu\nu}+R_{\mu\nu\rho\sigma}R^{\mu\nu\rho\sigma}.
\end{equation}
Here $R_{\mu\nu}$ is the Ricci curvature tensor. $S^{SC}$ is the string cloud action representing the cloud of strings elongated along the radial direction into the bulk from the boundary and is given by,
\begin{equation}
    S^{SC} = -\frac{1}{2}\sum_i T_i \int d^2\sigma  \sqrt{-\gamma} \gamma^{cd} \partial_c X^\mu \partial_d X^\nu g_{\mu\nu},
\end{equation}
where $\gamma^{cd}$ is the induced metric on the world sheet of the string, $T_i$ is the tension of the $i^{th}$ string hanging from the boundary to the horizon and ($c,\,d$) represents the world sheet coordinates.
The equation of motion obtained with respect to the bulk metric $g_{\mu\nu}$ from equation (\ref{eq_bulk_action}) can be written as,
\begin{equation}
    G_{\mu\nu} + \Lambda g_{\mu\nu} + \alpha H_{\mu\nu} = T_{\mu\nu}^{EM} + 8\pi G_5 T_{\mu\nu}^{SC}.
	\label{eq_EoM}
\end{equation}
Here,
\begin{equation}
    G_{\mu\nu} = R_{\mu\nu}-\frac{1}{2} g_{\mu\nu} R,
\end{equation}
is the Einstein tensor and 
\begin{equation}
    H_{\mu\nu} = 2 \left(R R_{\mu \nu } - 2 R_{\mu \zeta } R^{\zeta }{}_{\nu } - 2 R^{\alpha \eta } R_{\mu \nu \zeta \eta } +  R_{\mu }^{\zeta \eta \gamma }R_{\nu \zeta \eta \gamma } \right) - \frac{g_{\mu \nu } \hat{L}}{2},
\end{equation}
is the contribution from the GB term. The stress-energy tensor for Maxwell electro-magnetic(EM) contribution is,
\begin{equation}
    T_{\mu\nu}^{EM} = 2 F_\mu^\lambda F_{\nu\lambda} - \frac{1}{2}g_{\mu\nu} F^2,
\end{equation}
whose non-zero component is given as \cite{Pokhrel2023},
\begin{equation}
    F_{rt} = \sqrt{3} \frac{q}{r^3}.
\end{equation}
Here, $q$ is the integration constant related to the charge parameter of the black hole. The non-vanishing components of stress-energy tensor for the string cloud is given by \cite{Chakrabortty2011a,Pokhrel2023},
\begin{equation} 
    T_{tt}^{SC} = -\frac{a}{r^3}g_{tt},\hspace{1cm} T_{rr}^{SC} = -\frac{a}{r^3}g_{rr},
\end{equation}
where $a$ is the average string cloud density defined as \cite{Chakrabortty2011a},
\begin{equation}
	a(x) = \frac{1}{V_{3}}\int a(x) d^{3}x=\frac{T}{V_3} \sum_{i=1}^{N} \int \delta_{i}^{3} (x-X_i)d^3 x=\frac{T}{V_3}\sum_{i=1}^{N}1=\frac{T N}{V_3},
\end{equation}
where $V_3$ represents the $3$ dimensional space and $N$ is the number of strings. Now, considering $V_3$ along with $N$ going to infinity holding $N/V_3$ constant. Further, considering the static spherically symmetric metric ansatz for the bulk geometry as,
\begin{equation}
	ds^2 = -V(r) dt^2 + \frac{dr^2}{V(r)} + r^2 d\Omega^2_3,
\end{equation}
the $tt-$component of the equation of motion (\ref{eq_EoM}) is obtained as,
\begin{equation}
    \frac{V(r)}{2}\left[ -2\Lambda + \frac{12 \alpha V'(r) (V(r) -1)}{r^3} + \frac{6 (1-V(r))-3r V'(r)}{r^2}   \right] = 3\frac{q^2}{r^6} V(r) + 8\pi G_5 \frac{a}{r^3} V(r).
\end{equation}
From the above equation, the bulk metric solution $V(r)$ is calculated as,
\begin{equation}
    V(r) = 1 + \frac{r^2}{4\alpha}\left[  1 - \left( 1 + \frac{4}{3} \alpha \Lambda - 8 \frac{q^2 \alpha}{r^6} +  \frac{16 a \alpha}{3 r^3} + \frac{32 \alpha m}{r^4}   \right)^{\frac{1}{2}}   \right],
\end{equation}
where $8\pi G_5 = 1$, $m$ is the integration constant which is related to the mass of the black hole and the cosmological constant $\Lambda$ is related to the AdS radius $l$ as $\Lambda = -\frac{6}{l^2}$. Further, $V(r)$ becomes asymptotically AdS if the GB coupling belongs to the range $0\le \alpha \le\frac{l^2}{8}$.  By changing the coordinate $r=\frac{l^2}{u}$ and defining the boundary of the geometry at $u=0$, the metric solution is rewritten as,
\begin{equation}
V(u) =1+\frac{l^4}{4 \alpha  u^2}\left(1 -\sqrt{1+\frac{32 \alpha  m u^4}{l^8}-\frac{8 \alpha }{l^2}-\frac{8 \alpha  q^2 u^6}{l^{12}}+\frac{16 a \alpha  u^3}{3 l^6}}\right)=f(u) h(u),
	\label{eq_Vu}
\end{equation}
where $f(u)$ and $h(u)$ are given as, 
\begin{equation}
	f(u) = \frac{l^2}{u^2},
\end{equation}
and 
\begin{equation}
	h(u) = \frac{u^2}{l^2} + \frac{l^2}{4 \alpha}\left( 1-\sqrt{1 + \frac{32 \alpha m u^4}{l^8} - \frac{8\alpha}{l^2} - \frac{8 q^2 \alpha u^6}{ l^{12}}+ \frac{16 a \alpha u^3}{3 l^6}}  \right).
	\label{eq_hu}
\end{equation}
The mass parameter $m$ is calculated using the fact that $V(u=u_h)=0$ which gives,
\begin{equation}
	m = \frac{3 l^6 + 6 \alpha u_h^4 + l^4 u_h(3+4\Phi^2) - 2 a l^2 u_h^3}{12 u_h^4}.
\end{equation}
Here, baryon potential ($\Phi$) is related to the charge parameter $q$ as $ \Phi= \frac{\sqrt 3}{2}\frac{q u_h^2}{l^4}$. \\
The temperature of the black hole is calculated and it takes the form as,
\begin{equation}
	T = \frac{6l^4+3l^2 u_h^2-4 l^2 u_h^2 \Phi^2 - a u_h^3}{6 l^4 \pi u_h + 24 \pi \alpha u_h^3 }.
\end{equation}
The $\frac{\eta}{s}$ ratio of $4D$ charged black holes in Gauss–Bonnet gravity surrounded by a string cloud and quintessence is calculated in \cite{Sadeghi:2022kgi}, dropping the quintessence term, the equation reduces to,
\begin{equation}
	\frac{\eta}{s} = \frac{1}{4\pi}\left[1 - \frac{8 \alpha}{l^2} + \frac{16 \alpha \Phi^2}{3 l^4} u_h^2 -\frac{4 \alpha a}{3 l^6} u_h^3\right].
	\label{eq_eta_by_s}
\end{equation}
The ratio $\frac{\eta}{s}$ would be positive definite if the following constraint is satisfied,
\begin{equation}
	 4\alpha a u_h^3 - 16 l^2 \alpha \Phi^2 u_h^2 + 24 \alpha l^4 - 3 l^6 \leq 0.
	 \label{eq_constraint_eta_by_s}
\end{equation}
The discriminant of the above equation in terms of $u_h$ is given as,
\begin{equation}
	\Delta = - 48 l^8 (l^2-8\alpha) \alpha^2 \left[81 a^2 (l^2-8\alpha) + 1024 l^2 \alpha \Phi^6\right].
	\label{eq_discriminant_eta_by_s}
\end{equation}
The discriminant is negative definite for $\alpha < \frac{l^2}{8}$, which indicates that the constraint equation (\ref{eq_constraint_eta_by_s}) has only one real root for $u_h$. The single real root of equation (\ref{eq_constraint_eta_by_s}) corresponds to the single black hole solution region which can also be obtained from the expression of temperature as a function of horizon radius $u_h$ given by,
\begin{equation}
	(24 \pi \alpha T + a) u_h^3 + (4 l^2 \Phi^2- 3 l^2) u_h^2 + 6 l^4 \pi T u_h - 6 l^4=0.
	\label{eq_cubic_uh_temperature}
\end{equation}
From the above equation it is clear that there will be one real black hole solution when $\Phi\ge \frac{\sqrt{3}}{2}$ even at zero temperature. For the following analysis of transport properties, we have constrained the parameters as $0\le \alpha \le \frac{l^2}{8}$, $\Phi\ge \frac{\sqrt{3}}{2}$, $a\ge 0$, $T\ge0$ and $0\le v \le 1$, so that the viscosity to entropy density ratio equation (\ref{eq_eta_by_s}) is positive definite throughout the following analysis. When the magnitude of $\Phi$ is chosen slightly below $\frac{\sqrt{3}}{2}$, irrespective of the values of flavor density and GB coupling, the system admits three distinct black hole solutions and shows Hawking-Page like transition which has been discussed in \cite{Dey:2023inw}, thereby rendering the obtained observable results invalid.

With the above discussion we now proceed towards the study of transport properties for this set up in the following sections.

\section{Drag Force}
\label{sec_drag_force}
\begin{figure}[!ht]
\centering
\includegraphics[width=0.49\linewidth]{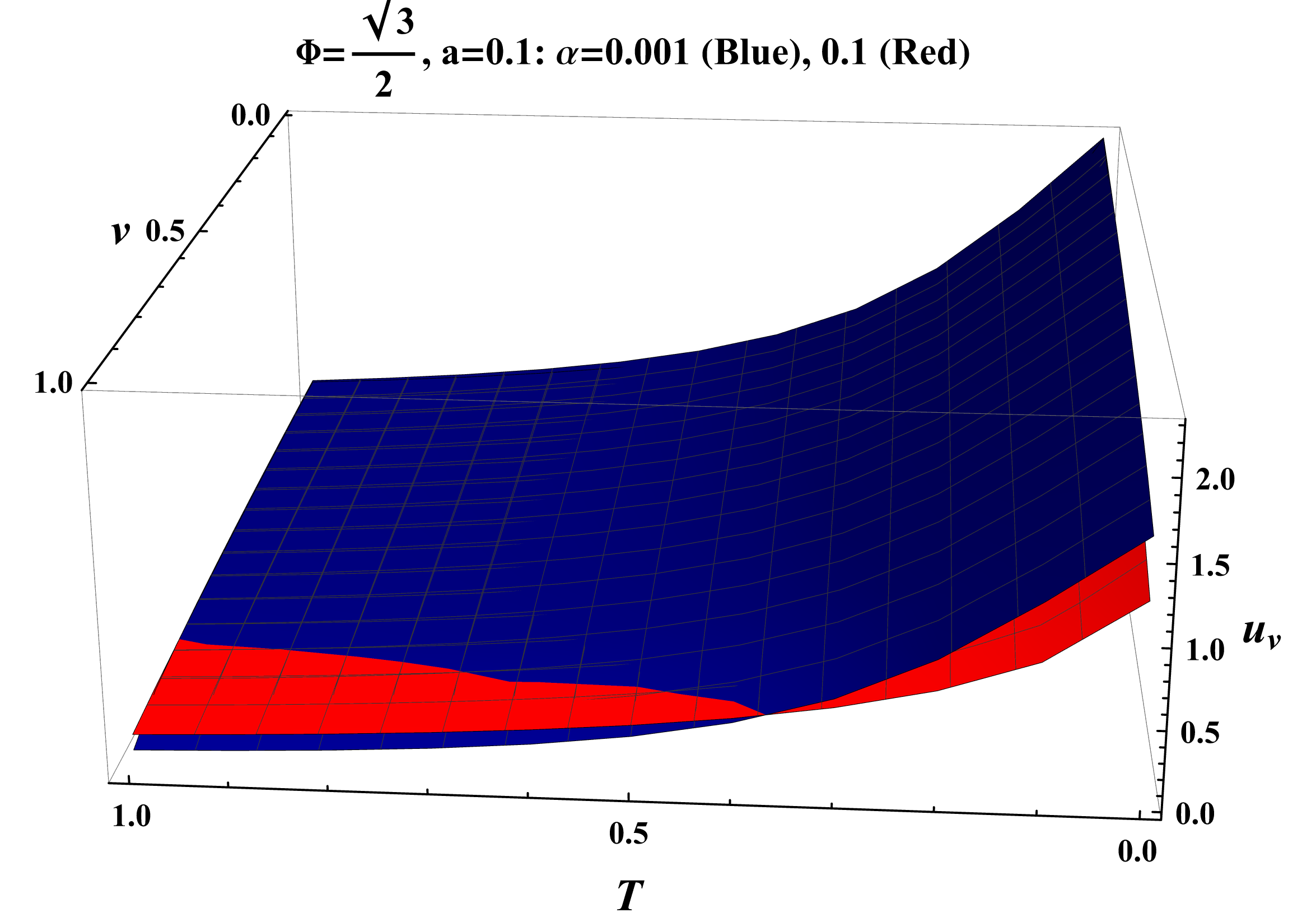}
\caption{Variation of critical point ($u_v$) vs temperature ($T$) vs velocity ($v$) for different values of GB coupling ($\alpha$).}
\label{Drag_CrossOver}
\end{figure}

In this section, we will compute the drag force experienced by an external probe quark translating through the baryon rich back-reacted thermal plasma with finite 't Hooft coupling correction. The motion of the probe string is described by the standard Nambu-Goto (NG) action given as,
\begin{equation}
	S(\mathcal {C}) = -\frac{1}{2\pi \alpha_t}\int d^2\sigma \sqrt{-\gamma},
	\label{Nambu_Goto_action}
\end{equation}
where $\alpha_t$ is related to the tension of the probe string. For the simplicity of the calculation, the static gauge $(\tau=t,\,\sigma=u)$ and the following boundary conditions are considered,
\begin{equation}
X^\mu(\tau, \sigma) = (t = \tau, u = \sigma, x = vt + \xi(\sigma), y = 0, z = 0),
\label{gaugefixing} 
\end{equation}
where $\xi(u)$ arises due to the trailing profile of the string. The motion of the probe string is considered to be along $x-$ direction of the boundary coordinates. Using the above conditions, the conventional NG action becomes,
\begin{equation}
        \begin{split}
            S(\mathcal {C}) = -\frac{1}{2\pi \alpha_t}\int dt du \left[\frac{f(u)^2}{h(u)}\left[h(u) + h(u)^2 \xi^{\prime 2}-v^2\right]\right]^{\frac{1}{2}},
            \label{equation_dissipative_force_nambu_goto_action_in_terms_of_xi_and_v}
            \end{split}
        \end{equation}
where ($\prime$) denotes the derivative with respect to radial coordinate $u$. Furthermore, the above equation (\ref{equation_dissipative_force_nambu_goto_action_in_terms_of_xi_and_v}) does not explicitly depend on $\xi$ so that, the corresponding conjugate momenta ($\Pi_\xi$) can be evaluated as,
\begin{equation}
        \xi^\prime = \frac{2 \pi \alpha_t \Pi_\xi}{f(u) h(u)}\sqrt{\frac{(h(u)-v^2)}{\left(h(u) - (2\pi \alpha_t)^2 \frac{\Pi_\xi^2}{f(u)^2}\right)}}.
    \end{equation}
Since $\xi$ should be real, the numerator and denominator inside the square root has to be zero at the same critical point along the radial coordinate called as $u_v$ which sets following two constraints,
\begin{equation}
        \begin{split}
       h(u_v) - v^2  = 0,\\
       h(u_v)f(u_v)^2 - (2\pi \alpha_t)^2 \Pi_\xi^2=0.
        \label{equation_dissipative_force_polynomial}
        \end{split} 
    \end{equation}
	Using the above two constraints, the conjugate momenta is obtained as,
	 \begin{equation}
        \Pi_\xi = \pm \frac{1}{2\pi \alpha_t} \frac{v l^2}{u_v^2}.
    \end{equation} 
	Since the drag force is defined as the rate of momentum loss by the probe during the interaction with the medium, which is the flow of total string momentum from the boundary to the horizon, the drag force is obtained  by considering the conservation of worldsheet current around a closed loop on the probe string which gives \cite{Gubser_2006},
	\begin{equation}
 F = -\frac{1}{2\pi \alpha_t}\frac{v l^2}{u_v^2}.
 \label{equation_dissipative_force_dissipative_force}
 \end{equation}
 Using the standard gauge/gravity relation,
 \begin{equation}
    \frac{l^4}{\alpha_t^2} = g^2_{YM} N=\lambda,
\end{equation}
where $g^2_{YM}$ and $N$ are the Yang-Mills gauge coupling and order of the gauge group respectively. The drag force equation (\ref{equation_dissipative_force_dissipative_force}) is now obtained as \cite{Pokhrel_2025,Pokhrel_2025a,Chakrabortty2016a},
\begin{equation}
	F =-\frac{\sqrt{\lambda}}{2\pi}  \left(\frac{v}{u_v^2(T,a,\alpha,\Phi)}\right),
	\label{eq_drag_force}
\end{equation}
where $v$ is the velocity of the probe quark, $\lambda$ is the 't Hooft coupling of the boundary gauge theory and $u_v$ is the solution of the equation,
\begin{equation}
	h(u_v) - v^2= \frac{u_v^2}{l^2} + \frac{l^2}{4 \alpha}\left( 1-\sqrt{1 + \frac{32 \alpha m u_v^4}{l^8} - \frac{8\alpha}{l^2} - \frac{8 q^2 \alpha u_v^6}{ l^{12}}+ \frac{16 a \alpha u_v^3}{3 l^6}}  \right)-v^2=0.
	\label{eq_huv_solution}
\end{equation}
Since, the above equation (\ref{eq_huv_solution}) is a sextic polynomial in $u_v$, its analytical solution is difficult to obtain. Hence, we have resorted to numerical methods to solve the equation and to compute the drag force. Further, while plotting we have rendered the parameters dimensionless by AdS radius $l$ as,
\begin{equation}
a\rightarrow \frac{a}{l},\,\alpha \rightarrow \frac{\alpha}{l^2},\,q\rightarrow \frac{q}{l^2},\,u_h \rightarrow \frac{u_h}{l},
\end{equation}
such that 
\begin{equation}\Phi(q,u_h) \rightarrow \Phi\left(\frac{q}{l^2},\frac{u_h}{l}\right),\,T(u_h, a, \alpha, \Phi(q,u_h))\rightarrow T\left(\frac{u_h}{l},\frac{a}{l},\frac{\alpha}{l^2},\Phi(\frac{q}{l^2},\frac{u_h}{l})\right),
\end{equation}
$l$ and $\lambda$ has been set to unity.
 The drag force calculated using equation (\ref{eq_drag_force}), as a function of different set of parameters is given in the following figure (\ref{fig_Drag_3D_Alphas}). 
\begin{figure}[!h]
	\centering
	\subfigure[]{\includegraphics[width=0.49\linewidth]{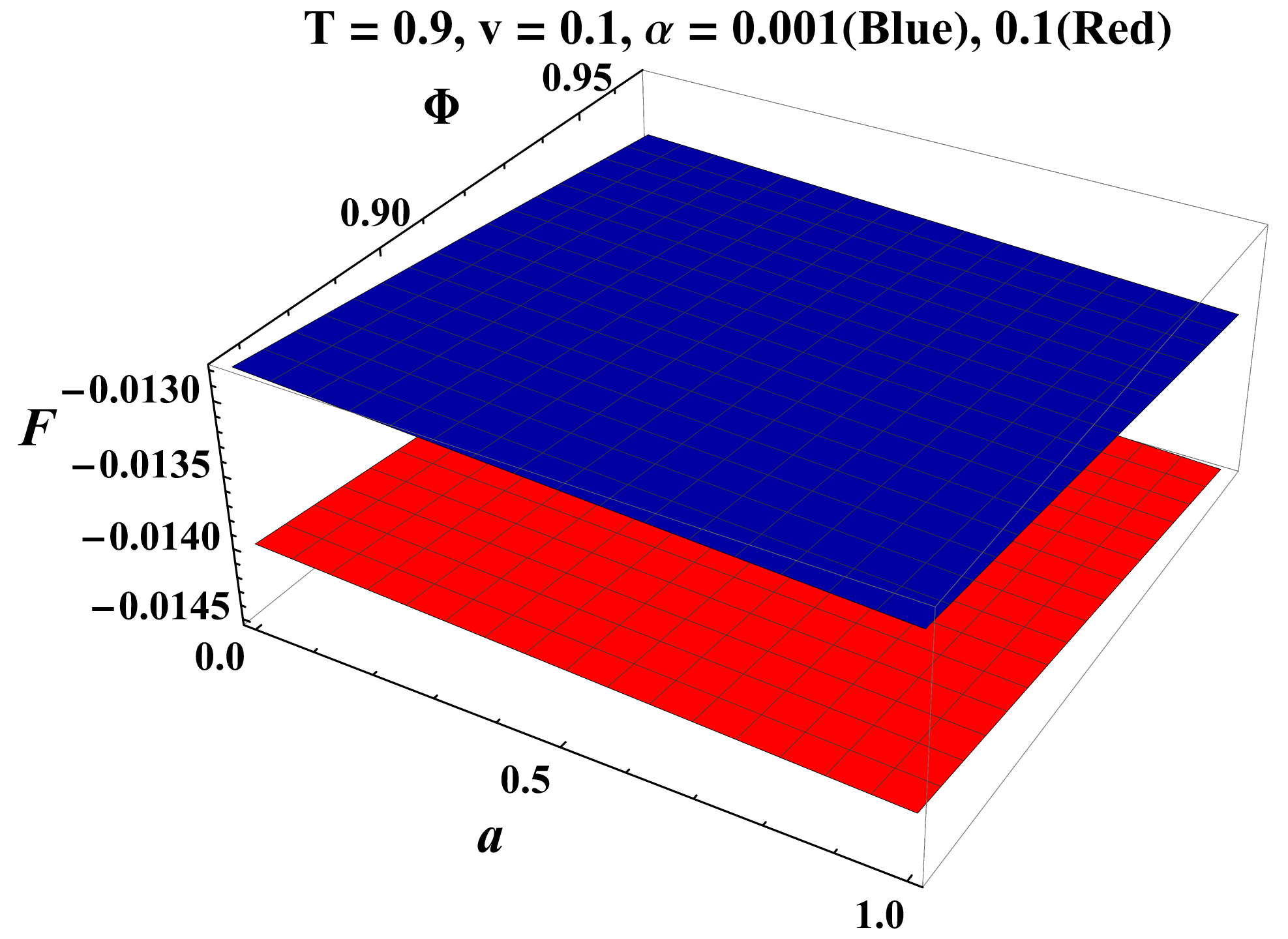}}
	\subfigure[]{\includegraphics[width=0.49\linewidth]{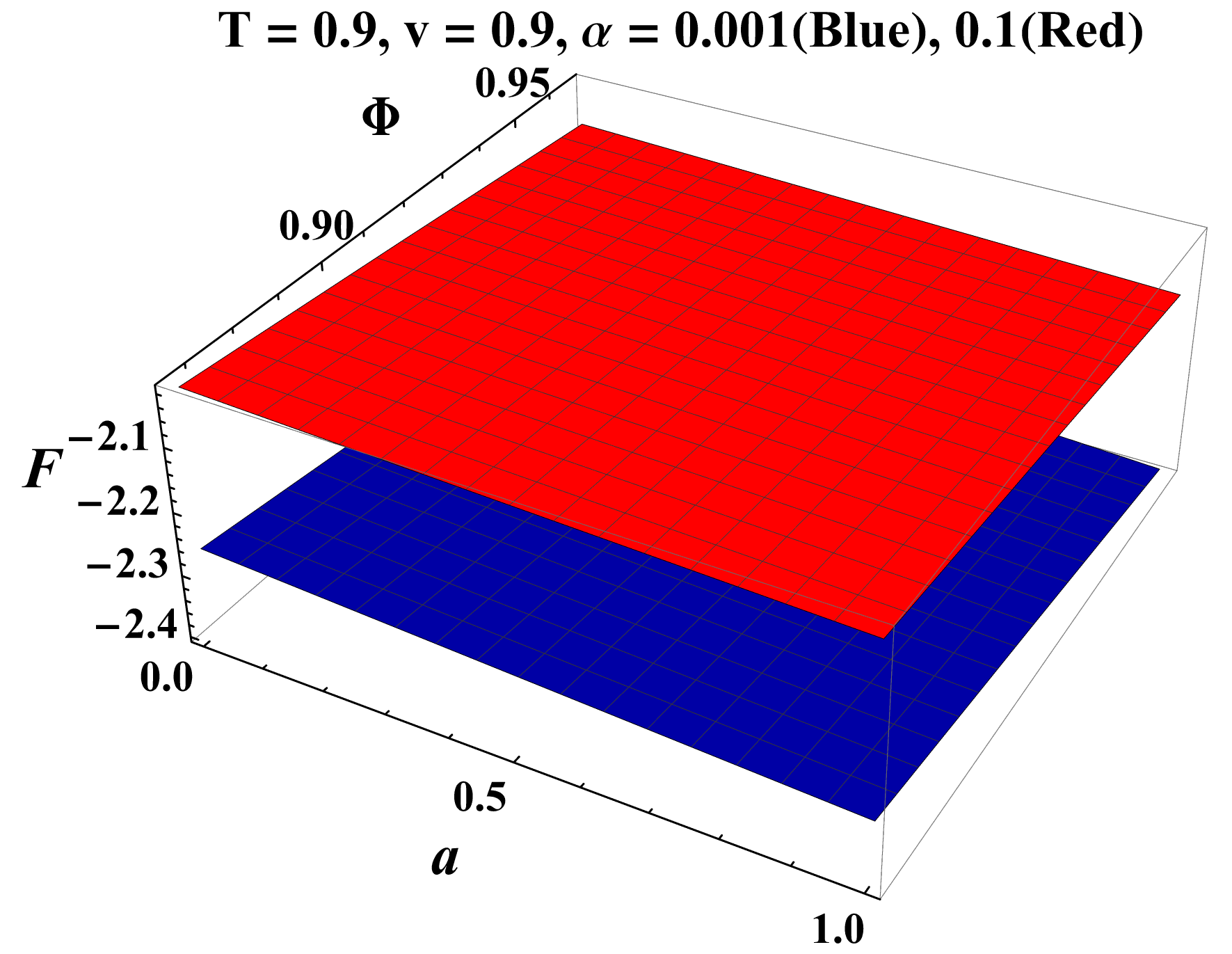}}
	\subfigure[]{\includegraphics[width=0.49\linewidth]{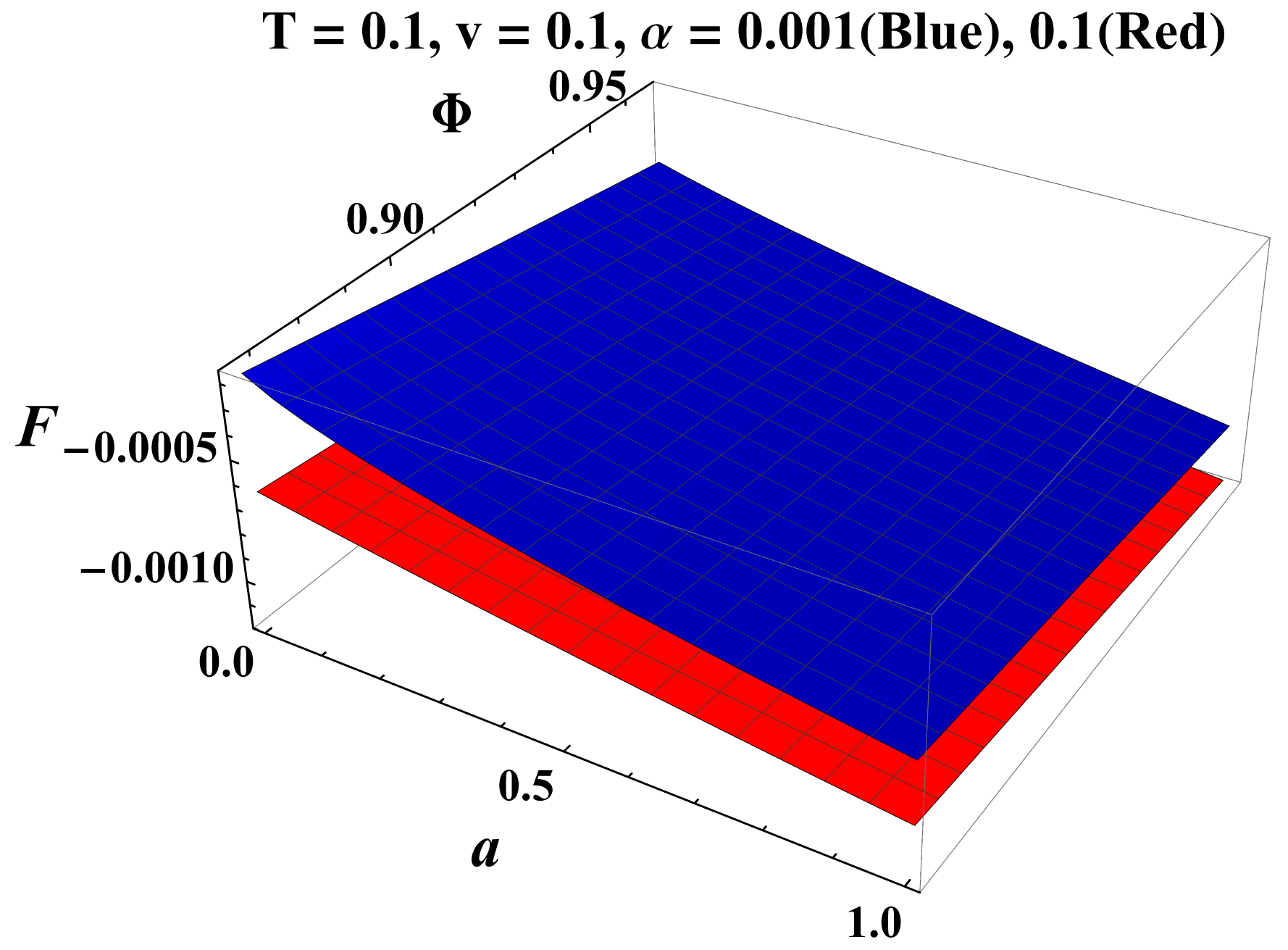}}
	\subfigure[]{\includegraphics[width=0.49\linewidth]{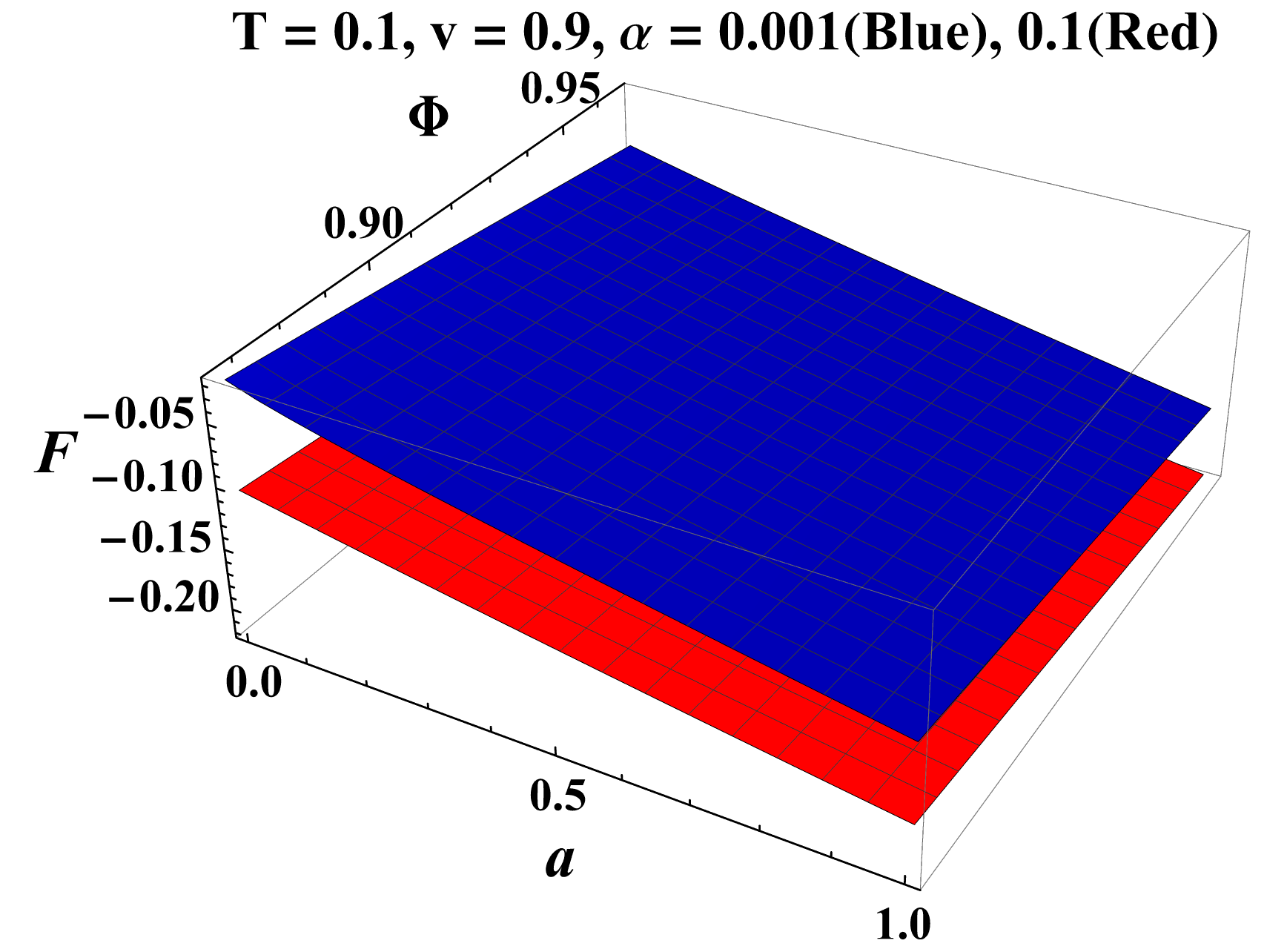}}
	\caption{Plot of drag force  vs flavor density and baryon potential for different values of temperature, velocity and GB coefficient.}
	\label{fig_Drag_3D_Alphas}
\end{figure}
In figure (\ref{fig_Drag_3D_Alphas}), the dependency of drag force has been analysed as a function of the parameters ($a, \alpha, \Phi$) for different probe velocity and temperature. It is observed that with the increase of temperature (bottom to top row), probe velocity (left to right column), flavor density and potential, the dissipative force experienced by the probe quark gets enhanced. It is evident from figure (\ref{Drag_CrossOver}) that the critical point ($u_v$) shifts toward lower values with increasing temperature and velocity due to more trailing profile of the string. Moreover, the decline in $u_v$ occurs more rapidly for smaller values of the GB coupling $\alpha$ than for larger value of it since deformation of AdS spacetime is more. As a result, the drag force becomes slightly weaker at higher temperature and velocity for larger $\alpha$, than the smaller value of it which is depicted in figure (\ref{fig_Drag_3D_Alphas} b).

\section{Jet Quenching Parameter}
\label{sec_jetQ}
\begin{figure}[!h]
	\centering
	\subfigure[]{\includegraphics[width=0.45\linewidth]{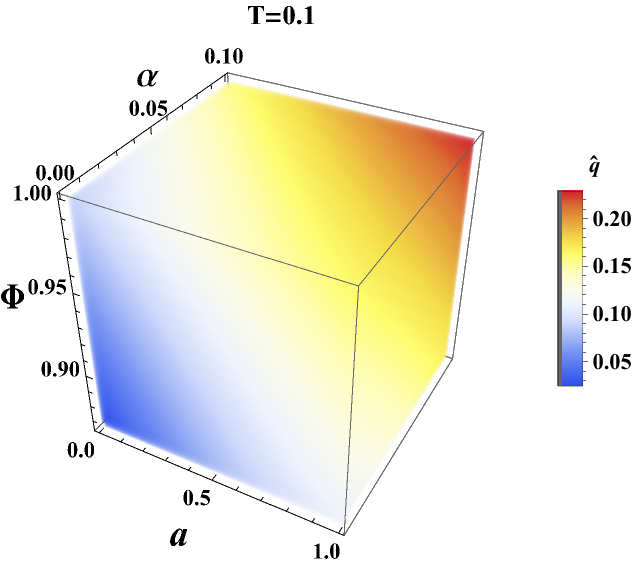}}
	\subfigure[]{\includegraphics[width=0.45\linewidth]{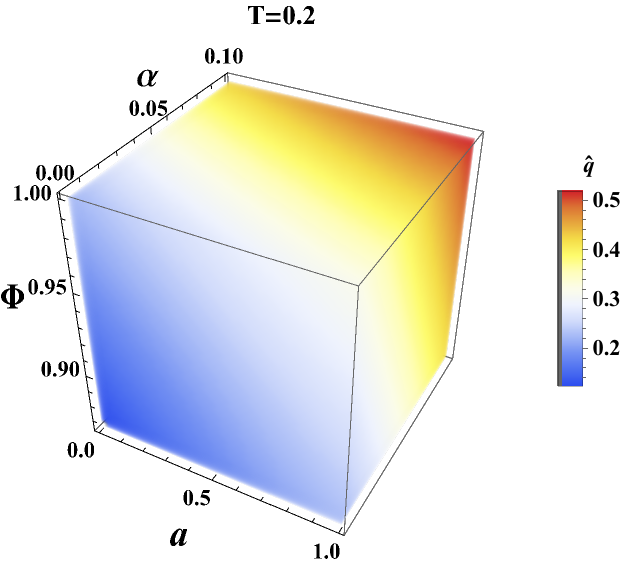}}
	\caption{Plot of flavor density vs baryon potential vs GB coefficient with volume plot representing jet quenching parameter ($\hat{q})$ for (a) $T=0.1$. (b) $T=0.2$.}
	\label{fig_jetQ_3D_plots}
\end{figure}

In this section, we compute the jet quenching parameter ($\hat{q}$) using holographic approach \cite{Liu:2006ug}. The $\hat{q}$ is related with the expectation value of the Wilson loop traced by $q\bar{q}$ pair with separation length L and the latter one is connected to the regularised Nambu-Goto action. The relation between the jet quenching parameter and the expectation value of light-like Wilson loop in the adjoint representation is given as \cite{Liu:2006he}
\begin{equation}
	\left< W^A(\mathcal{C})\right> = e^{-\frac{1}{4\sqrt{2}}\hat{q} L^- L^2},
	\label{equation_jet_quenching_Wilson_loop_expect_adjoint}
\end{equation}
where $\mathcal{C}$ is traced out by the $q\bar{q}$ separation length $L$ and length $L^-$ along the light cone of the boundary gauge theory. Using the gauge/gravity duality, we can compute the expectation value of the Wilson loop in the fundamental representation \cite{Liu:2006he,Maldacena1998},
\begin{equation}
	\left< W^F (\mathcal{C})\right> = e^{i S(\mathcal{C})}.
	\label{equation_jet_quenching_Wilson_loop_expect_funda_Nambu_goto}
\end{equation}
Where, $S(\mathcal{C})$ is the Nambu-Goto action of the fundamental probe string whose two end points are attached to the boundary corresponding to the $q\bar{q}$ pair. The relation between the expectation value of fundamental representation and adjoint representation of Wilson loop can be constructed and it takes the simple form as,
 \begin{equation}
	\left< W^F(\mathcal{C})\right>^2 = \left< W^A(\mathcal{C})\right>.
	\label{equation_jet_quenching_Wilson_loop_expect_funda_adjoint_relation}
 \end{equation}
Now from equations (\ref{equation_jet_quenching_Wilson_loop_expect_adjoint}, \ref{equation_jet_quenching_Wilson_loop_expect_funda_Nambu_goto}) and (\ref{equation_jet_quenching_Wilson_loop_expect_funda_adjoint_relation}), the jet quenching parameter ($\hat{q}$) is defined as,
\begin{equation}
	\hat{q} = -\frac{8 \sqrt{2}i}{L^- L^2}(S-S_0).
\end{equation}
Here $S_0$ is the action for the self energy contribution of the total mass of $q\bar{q}$ pair. Considering the light-cone coordinate the spacetime metric solution takes the form as,
\begin{equation}
	ds^2 = f(u)\left[-(1+h(u)) dx^+ dx^- + \frac{1}{2}(1-h(u)) \{dx^{+2}+dx^{-2}\}+dy^2 + dz^2 + \frac{du^2}{h(u)}\right].
	\label{equation_jet_quenching_metric_fu_hu}
\end{equation}
Here, $x^\pm$ is the light-cone coordinate defined as,
\begin{equation}
	x^\pm = \frac{t \pm x}{\sqrt{2}}.
\end{equation}
With the choice of static gauge as $\tau = x^- (L^-\ge x^- \ge 0)$, $ \sigma = y (-\frac{L}{2}\le y \le \frac{L}{2} )$, $q\bar{q}$ pair at $y=\pm \frac{L}{2}$,  $x^+ = $ constant and $z = $ constant plane, the Nambu-Goto action of the probe string is obtained as,
\begin{equation}
	S = \frac{i L^-}{\sqrt{2} \pi \alpha_t}\int_{0}^{\frac{L}{2}} dy f\sqrt{(1-h) (1+\frac{u'^2}{h} )}.
\label{langrangian}
\end{equation}
Following the standard approach \cite{Liu:2006ug}, the equation of motion for the probe string and the solution for $u'$ is obtained. Finally, regularisation of the Nambu-Goto action by subtracting the self energy contribution gives,
\begin{equation}
	S -S_0 \approx \frac{i L^- L^2}{8 \sqrt{2} \pi \alpha_t I_1}.
\end{equation}
 Using these relation, the $\hat{q}$ can be expressed as \cite{Pokhrel_2025,Pokhrel_2025a,Chakrabortty2016a},
\begin{equation}
	\hat{q} = \frac{\sqrt{g^2_{YM}N}}{\pi I_1(T,a,\alpha,\Phi)},
	\label{eq_jetq_formula}
\end{equation}
where $N$ is the order of the gauge group, $g_{YM}$ is Yang-Mills gauge coupling and
\begin{equation}
	I_1 = \int_\delta^{u_h(T,a,\alpha,\Phi)} \frac{u^2 du}{\sqrt{h(h-1)}}.
	\label{eq_jetq_I1_formula}
\end{equation}
Here, $\delta$ is the cut-off imposed to keep away from the imaginary value of $u^\prime$, the derivative of radial direction with respect to $\sigma$ which is space-like coordinate along the string. Finally, after completion of the calculation the cut-off is set to zero.
The nature of $\hat{q}$ considering the metric solution (\ref{eq_hu}) is plotted against parameters $a,\alpha,\Phi$ for different temperature in the following plot (\ref{fig_jetQ_3D_plots}).
From figure (\ref{fig_jetQ_3D_plots}), it is concluded that increase of temperature, flavor density, GB coefficient and baryon potential leads to a rise in the value of $\hat{q}$. This indicates that the energy loss of the probe quark due to the suppression of the high transverse momentum in the thermal medium increases with enhancement of these parameters. 
\section{Screening Length}
\label{sec_screening_length}
This section deals with the study of the screening length of quark-antiquark pair. The screening length is defined as the stable maximum separation between the pair beyond which they get separated into two individual parts as unbound state configuration. The $q\bar{q}$ pair is considered through an external probe string having its both ends attached to the boundary and its body hanging into the radial bulk direction with a turning point in between. We study the screening length of the $q\bar{q}$ pair in both perpendicular and parallel orientation \cite{Zhang_2024,Zhu_2024,Pokhrel_2025,Pokhrel_2025a}. Here the orientations are defined by the alignment of the axis of the $q\bar q$ pair compare to the direction of motion. To study the screening length, the boost in the dual gravity is introduced as below,
\begin{equation}
	\begin{split}
	dt= cosh(\beta) dt^* - sinh(\beta) dz^*,\\
	dz = -sinh(\beta) dt^* + cosh (\beta) dz^*,
	\end{split}
\end{equation}
which recasts the metric as,
\begin{equation}
	\begin{split}
	ds^2 = f\left[-\{1-cosh^2(\beta) (1-h)\}dt^{*2} + \{1+ (1-h) sinh^2(\beta)\}dz^{*2}\right. \\ \left.-2(1-h) cosh(\beta) sinh(\beta) dt^* dz^* + dx^2 + dy^2 + \frac{du^2}{h}\right],
	\label{G_Metric_Screening_Length}
	\end{split}
\end{equation}
where $\beta = tanh^{-1} (v)$ is the rapidity parameter representing the pair's relative velocity with thermal plasma. In the following subsections, screening length for the perpendicular and parallel orientation are studied.  
\subsection{Perpendicular Orientation:}
\begin{figure}[!h]
	\centering
	\subfigure[]{\includegraphics[width=0.3\linewidth]{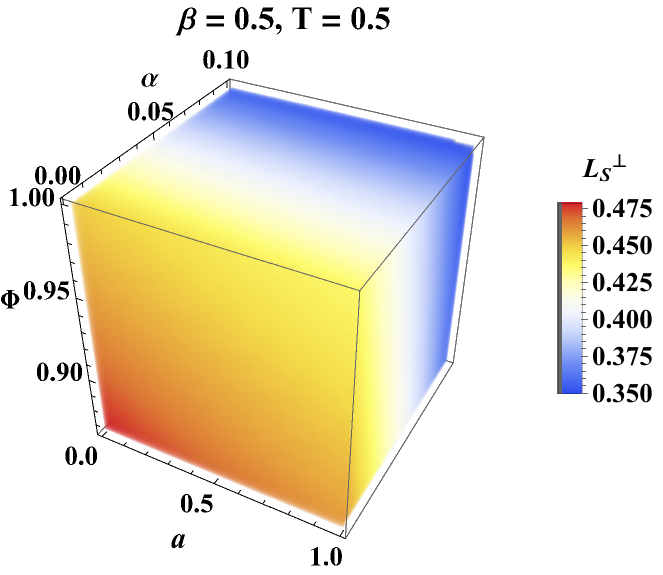}}
	\subfigure[]{\includegraphics[width=0.3\linewidth]{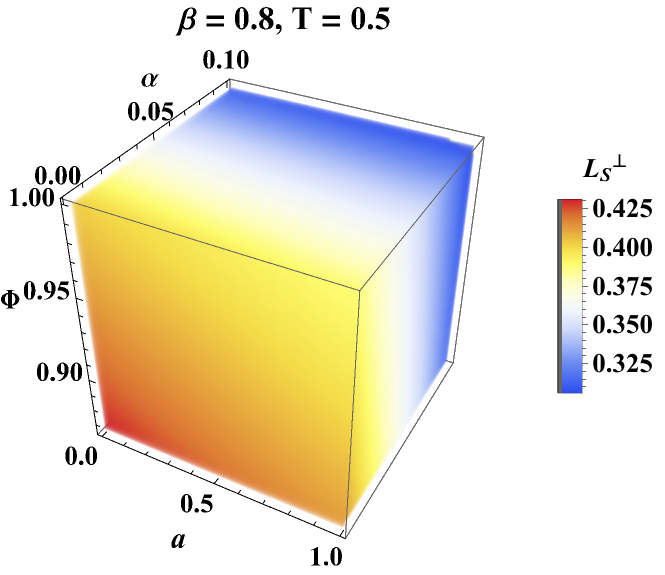}}
	\noindent\subfigure[]{\includegraphics[width=0.3\linewidth]{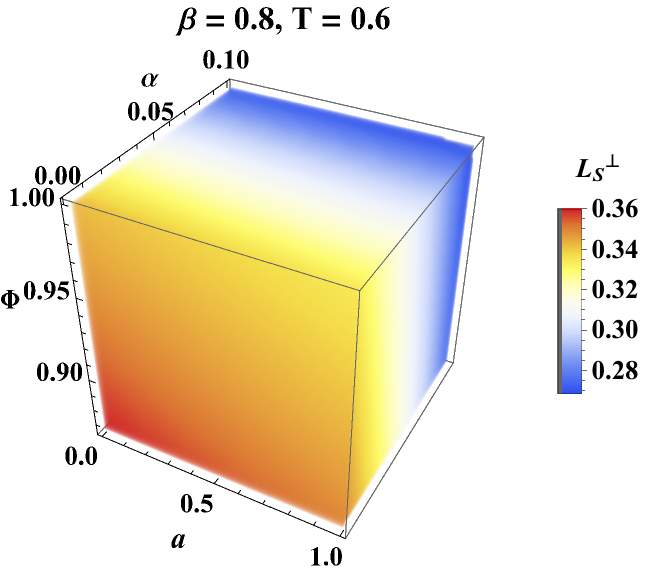}}
	\caption{Flavor density vs potential vs GB coupling with volume representing the perpendicular screening length $L_S^{\perp}$ for (a) $\beta=0.5,\,T=0.5$ (b) $\beta=0.8,\,T=0.5$ (c) $\beta=0.8,\,T=0.6$.}
	\label{fig_L_vs_W_perp_case1}
\end{figure}

For the choice of static gauge, time coordinate of the probe string world-sheet $\tau = t^*,\, \sigma = x,\, y=z^*=0$ and boundary condition on the probe string as,
\begin{equation}
	u(\sigma =\pm \frac{L}{2}) = 0,\, u(\sigma =0) = u_{ext},\, u'(\sigma=0) =0,
\end{equation}
the probe string action is obtained as,
\begin{equation}
	S = -\frac{\mathcal{T}}{2 \pi \alpha_t} \int d\sigma \sqrt{f^2[1 + (h-1) cosh^2(\beta)] + \frac{f^2}{h}[1 + (h-1) cosh^2(\beta)]u'^{2}}.
\label{wsheeta}
\end{equation}
Using the Hamilton's like equation since the above action does not explicitly depend on $\sigma$, the constant of motion $W$ is obtained as,
\begin{equation}
	\frac{\partial \mathcal{L}}{\partial u'} u' - \mathcal{L} = W.
\label{hequation}
\end{equation}
Using equations. (\ref{wsheeta}) and (\ref{hequation}), the solution for $u'$ is obtained as,
\begin{equation}
	u'^2 =  \frac{h[f^2 + f^2(h-1) cosh^2(\beta) -{W}^2]}{{W}^2}.
	\label{equation_screening_length_u_dashed_squared}
\end{equation}
The turning point of the probe string is obtained by setting $u' = 0$, which becomes true when $\left.h\right|_{u_{ext1}}=0$ or $\left.(f^2 + f^2(h-1) cosh^2 (\beta) - {W}^2 ) \right|_{u_{ext2}}=0 \label{ucsol}$. The first condition corresponds to the turning point at the horizon under which the tip of the probe string reaches the horizon and breaks down into two separate strings. The second condition corresponds to the turning point outside the horizon along the radial direction. The maximum separation length is obtained by considering the second condition as the turning point of the probe string. Hence,
the perpendicular separation length ($L^{\perp}$) is obtained after integrating Eq (\ref{equation_screening_length_u_dashed_squared}) and using the boundary condition $u(\sigma =\pm \frac{L}{2}) = 0$ as \cite{Pokhrel_2025},
\begin{equation}
	L^{\perp} = \int_{0}^{u_c} \frac{2 W du}{\sqrt{h}\sqrt{f^2(1+(h-1)cosh^2(\beta))-{W}^2}}.
	\label{screening_length_final_equation}
\end{equation}
 From equation (\ref{screening_length_final_equation}), the separation length is calculated and the maximum value of the separation distance called screening length ($L_S^\perp$) between the $q\bar{q}$ pair is extracted for a range of constant of motion $W$ of the probe string, flavor density $a$, GB coupling $\alpha$ and potential $\Phi$. It is shown by density plot in figure (\ref{fig_L_vs_W_perp_case1}). The increment in the rapidity parameter (subfigure a to b), temperature (subfigure b to c), flavor density, GB coupling and potential diminishes the screening length.
\subsection{Parallel Orientation:}
In this subsection, the separation distance between the $q\bar{q}$ pair in parallel orientation has been studied. With static gauge $\tau = t^*,\, \sigma = z,\, x=y=0,$ and boundary condition on the probe string as,
\begin{equation}
	u(\sigma =\pm \frac{L}{2}) = 0,\, u(\sigma =0) = u_{ext},\, u'(\sigma=0) =0,
    \label{equation_screening_length_parallel_boundary_conditions}
\end{equation}
the probe string action takes the form,
\begin{equation}
    S = -\frac{\mathcal{T}}{2\pi \alpha_t}\int d\sigma f\sqrt{h + \left[\frac{1}{h}+\left(1-\frac{1}{h}\right)cosh^2(\beta)\right]u'^2}.
\end{equation}
Similar to the perpendicular section, after setting the Hamilton's like equation we get,
\begin{equation}
    u'^2 = \frac{h(f^2h - {W} ^2)}{[1 + (h-1) cosh^2(\beta)]{W}^2}.
    \label{equation_screening_length_parallel_u_dashed_squared}
\end{equation}
From which the turning point ($u_c$) of the probe string is obtained by setting,
\begin{equation}
    \left. f^2 h - {W} ^2\right|_{u_{ext2}} = 0, 
\end{equation}
which is then integrated along with the boundary condition equation (\ref{equation_screening_length_parallel_boundary_conditions}), from which the parallel separation length ($L^{\parallel}$) is calculated as \cite{Pokhrel_2025},
\begin{equation}
    L^\parallel = \int_{0}^{u_c} \frac{2 W  \sqrt{1 + (h-1) cosh^2(\beta)}}{h\sqrt{f^2 h - {W}^2}}.
	\label{eq_L_parallel}
\end{equation}
Following similar approach as that of perpendicular case, we obtain the screening length for the parallel case from the above equation (\ref{eq_L_parallel}).
\begin{figure}[!h]
	\centering
	\subfigure[]{\includegraphics[width=0.3\linewidth]{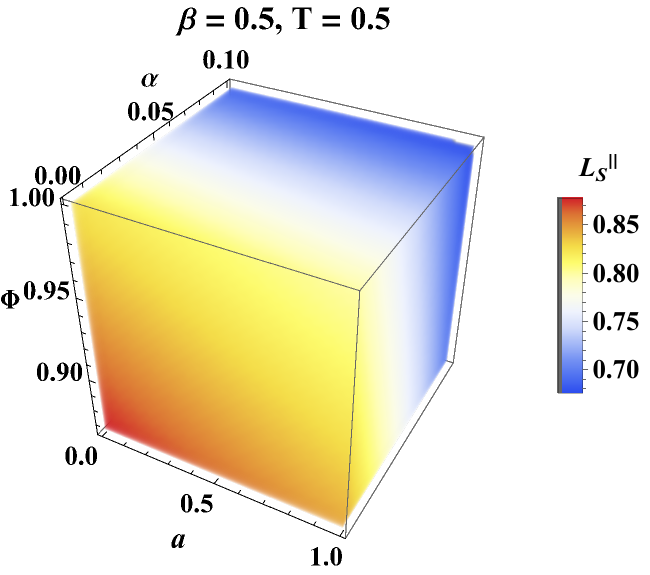}}
	\subfigure[]{\includegraphics[width=0.3\linewidth]{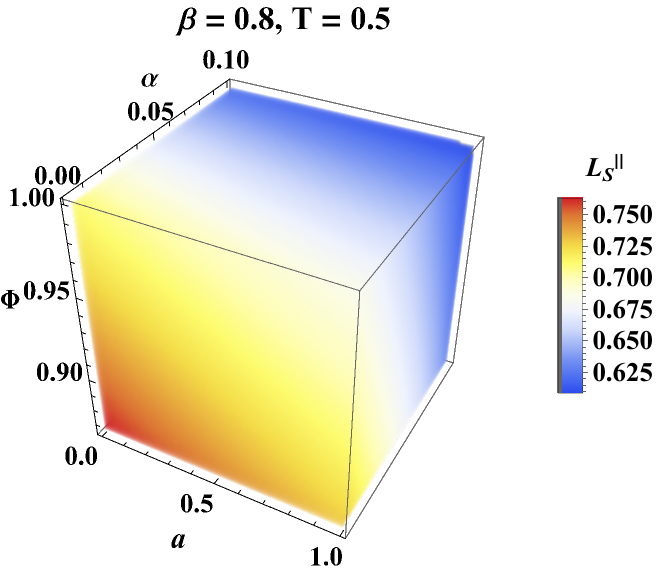}}
	\subfigure[]{\includegraphics[width=0.3\linewidth]{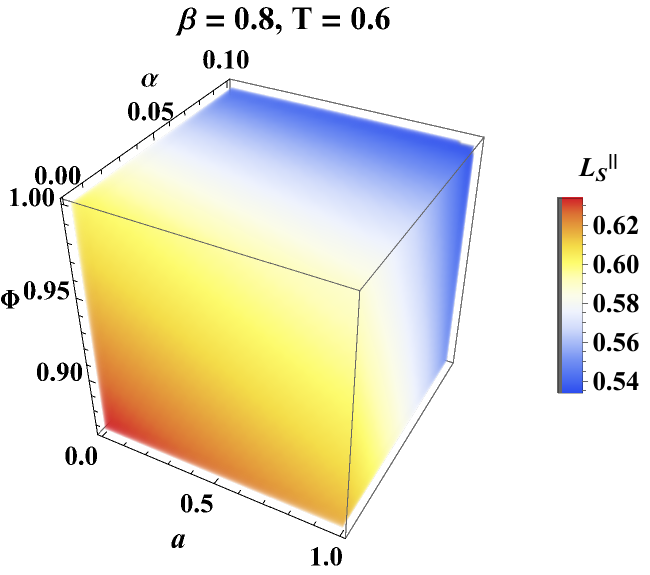}}
	\caption{Flavor density vs potential vs GB coupling with volume representing the parallel screening length $L_S^{||}$ for (a) $\beta=0.5,\,T=0.5$ (b) $\beta=0.8,\,T=0.5$ (c) $\beta=0.8,\,T=0.6$.}
	\label{fig_L_vs_W_para_case1}
\end{figure}
Similar conclusion is drawn as that of perpendicular case from figure (\ref{fig_L_vs_W_para_case1}). With increase in rapidity parameter (subfigure a to b), temperature (subfigure b to c), flavor density, GB coupling and potential the screening length gets reduced.

Further, comparing both the orientations, it is observed that the screening length in parallel orientation is greater than that of perpendicular orientation for the same set of parameters.
\section{Radial Profile of the Rotating string}
\label{sec_radial_profile}
\begin{figure}[!h]
	\centering
	\subfigure[]{\includegraphics[width=0.31\linewidth]{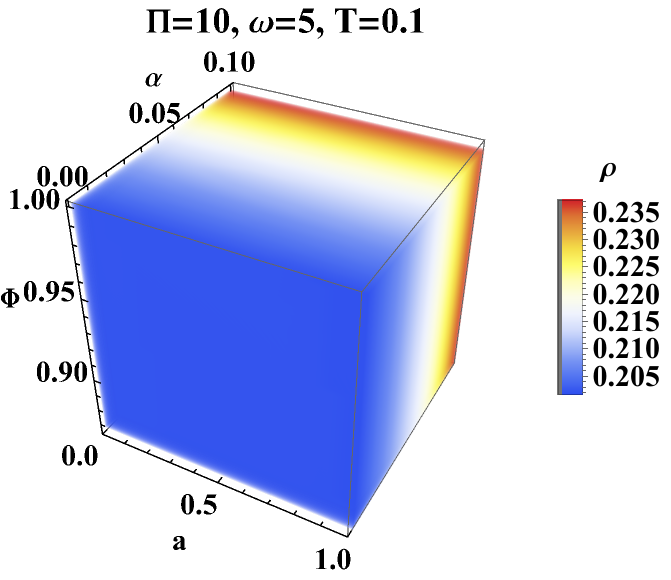}}\hspace{1cm}
	\subfigure[]{\includegraphics[width=0.31\linewidth]{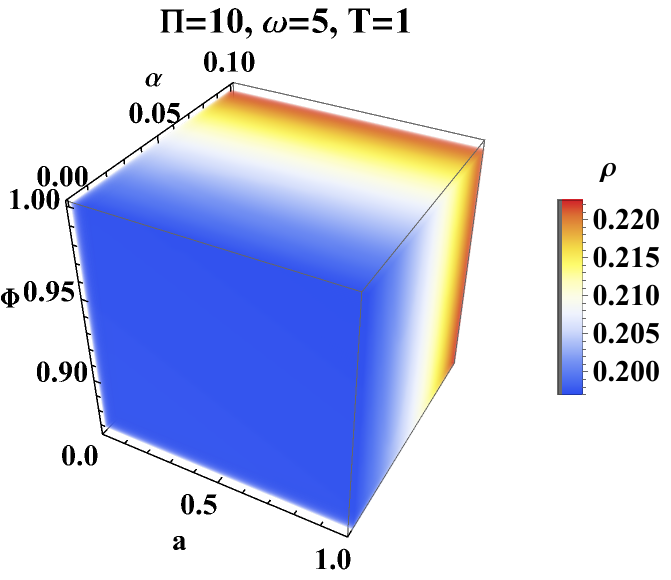}}
	\subfigure[]{\includegraphics[width=0.31\linewidth]{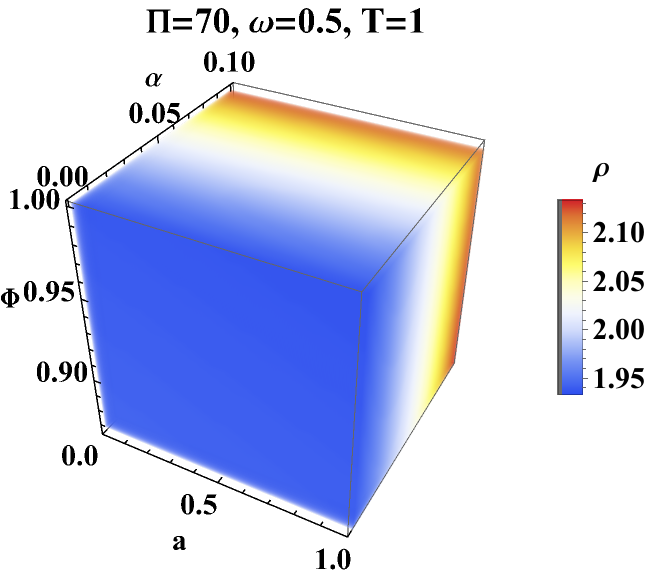}}\hspace{1cm}
	\subfigure[]{\includegraphics[width=0.31\linewidth]{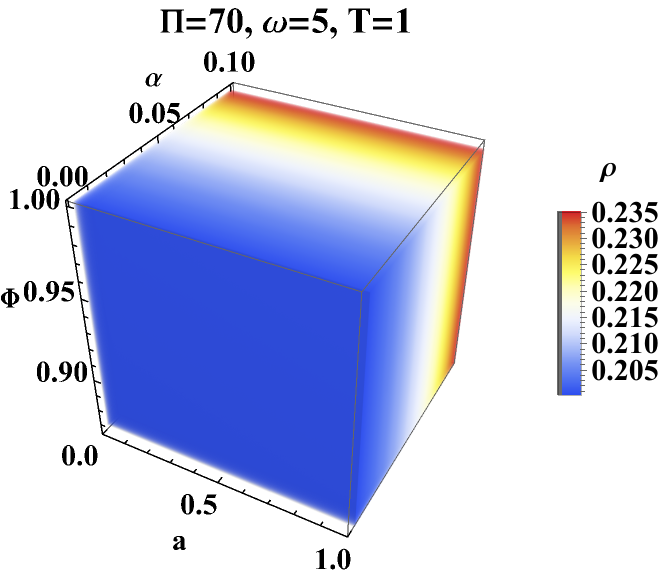}}
	\subfigure[]{\includegraphics[width=0.31\linewidth]{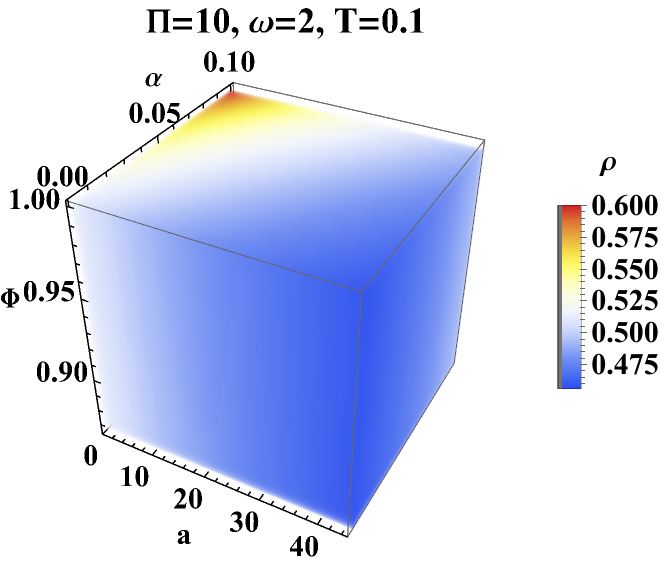}}
	
	\caption{ Flavor density vs potential vs GB coupling with volume representing the radial profile $\rho$ for different combination of parameters.}
	\label{fig_radial_b1}
\end{figure}

The radial profile of a constantly rotating probe string has been analysed in this section.
The radial profile is obtained by considering a probe string whose one end is attached to the boundary of the spacetime, which corresponds to the heavy probe quark in the dual gauge theory and rotating with constant angular frequency $\omega$ and radius $\mathcal{R}$ in two dimensional plane of the boundary. The velocity and acceleration are given as $v= \mathcal{R} \omega$ and $a= \omega^2 \mathcal{R}$. Following the approach prescribed in \cite{Pokhrel_2025,Pokhrel_2025a,Chakrabortty2016a}, we parameterize the string world sheet as,
\begin{equation}
	X^\mu(\tau,\sigma) = (t=\tau, u = \sigma, x = \rho(\sigma)cos(\omega t + \theta(\sigma)), y = \rho(\sigma)sin(\omega t + \theta(\sigma)), z=0),
	\label{equation_energyloss_parameterization}
\end{equation}
where $\rho(\sigma)$ represents the radial profile and $\theta(\sigma)$ represents the angular profile with the boundary conditions,
\begin{equation}
	\rho(0) = \mathcal{R}, \, \theta(0) = 0.
\end{equation}
The Lagrangian density is obtained for the standard NG probe string action as,
\begin{equation}
	\mathcal{L} = \left[\frac{f^2}{h}(h - \rho^2 \omega^2) + f^2 (h- \rho^2 \omega^2) \rho^{'2} + h f^2 \rho^2 \theta^{'2}\right]^{\frac{1}{2}}.
	\label{equation_energyloss_lagrangian_density}
\end{equation}
The above Lagrangian does not explicitly depends on $\theta$. Hence, the corresponding conjugate momenta is obtained as,
\begin{equation}
	\Pi_\theta  = \frac{\partial \mathcal{L}}{\partial \theta'} = \frac{h f^2 \rho^2 \theta'}{\mathcal{L}}.
\label{paitheta}
\end{equation}
The above equation is solved for $\theta'$ which gives,
\begin{equation}
	\theta' = \Pi_\theta \sqrt{\frac{(h- \rho^2 \omega^2)(1 + h \rho^{'2})}{h^2 \rho^2 (h f^2 \rho^2 - \Pi_\theta^2)}}.
	\label{equation_energyloss_theta_dashed}
\end{equation}
The factor $h-\rho^2 \omega^2$ in the above equation must changes sign at some critical point $u_c<u_h$, since this factor becomes negative at $u=u_h$ and is positive at the boundary. Further, the string in the range $0\le u<u_c$ are causally disconnected from the string in the range $u_c\le u \le u_h$, since $v^2=\rho^2\omega^2$ is less than unity in the former case which is the physically relevant part and $v^2$ is greater than unity in the latter. The probe string moving with local speed of light is represented by the curve $h(u_c)=\rho(u_c)^2 \omega^2$, which is intersected by all other probe string moving with speed slower than the speed of light at the critical point $u=u_c$. In order to have $\theta'$ to be real following constraints are imposed at the critical point,
\begin{equation}
	h(u_c) - \rho(u_c)^2 \omega^2 =0.
	\label{equation_energyloss_condition1}
\end{equation}
Use equation (\ref{equation_energyloss_theta_dashed}) to eliminate $\theta'$ from the equation of motion for $\rho$ coordinate which is calculated as, 
\begin{equation}
	\frac{\partial}{\partial u}\left(\frac{\partial \mathcal{L}}{\partial \rho'}\right)- \frac{\partial \mathcal{L}}{\partial \rho}=0.
\end{equation} 
 The equation of motion for $\rho$ is obtained as,
 \begin{align}
	& 2(\Pi_\theta^2 - \omega^2 f^2 \rho^4) - \left[2 f h \rho^3 (h -\omega^2 \rho^2) f' - \rho \{\Pi_\theta^2 - f^2 ( 2 h \rho^2 - \omega^2 \rho^4)\}h'\right]\rho'\nonumber \\ & + 2h (\Pi_\theta^2 - \omega^2 f^2 \rho^4 ) \rho^{'2} - \rho^3 \left[ 2 f h^2 (h- \omega^2 \rho^2) f' - (\Pi_\theta^2 \omega^2 - f^2 h^2 )h'\right]\rho^{'3} \nonumber \\ & + 2 \rho(h-\omega^2 \rho^2) (\Pi_\theta^2 - f^2 h \rho^2) \rho^{''} = 0.
	\label{equation_energyloss_final_EOM_for_rho}
\end{align} 
To obtain the radial profile, we fix the values of $\rho$ and $\rho'$ at the critical point $u_c$ for some fixed values of the parameters. The radial coordinate is expanded around the critical point and linear order terms are considered, which sets the equation for $\rho'(u_c)=\rho'_c$ as,
\begin{align}
	\rho'_c = &\frac{1}{(4 u_c h_c^3 \rho_c)}\left[- 4 u_c^2 h_c^2 - h_c^3 \rho_c^2 + \omega^2 h_c^2 \rho_c^4 + 2 u_c h_c^2 \rho_c^2 h'_c- \Pi^2 u_c^4 h^{'2}_c\right.\nonumber \\ & \left. + \left(16 u_c^2 h_c^5 \rho_c^2 + \{ h_c^3 \rho_c^2 + \Pi^2 u_c^4 h^{'2}_c + h_c^2 ( 4 u_c^2 - \omega^2 \rho_c^4 - 2 u_c \rho_c^2 h'_c)\}^2\right)^{\frac{1}{2}}\right].
\end{align}
 We have plotted radial profile with respect to the parameters $a$, $\alpha$ and $\Phi$ in figure (\ref{fig_radial_b1}) in the form of a density plot.
From figure (\ref{fig_radial_b1}), it is observed that the radial profile is reduced with increase in temperature (subfigure a to b), angular frequency (subfigure c to d) and is enhanced with conjugate momenta (subfigure b to d). Furthermore, figure (\ref{fig_radial_b1}e) confirms that the radial profile gets reduced with increase in flavor density and baryon potential, whereas it gets enhanced with increase in GB coupling.

\section{Energy Loss of the Rotating String}\label{sec_energy_loss}

\begin{figure}[!h]
	\centering 
	\subfigure[]{\includegraphics[width=0.32\linewidth]{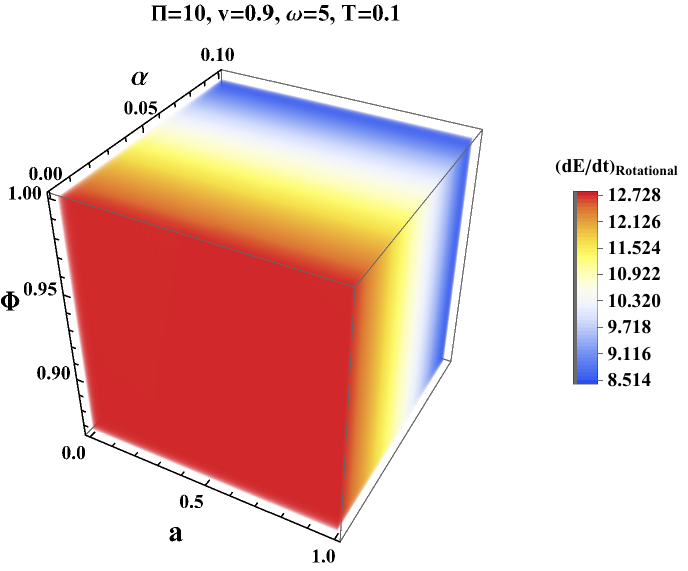}}
	\subfigure[]{\includegraphics[width=0.32\linewidth]{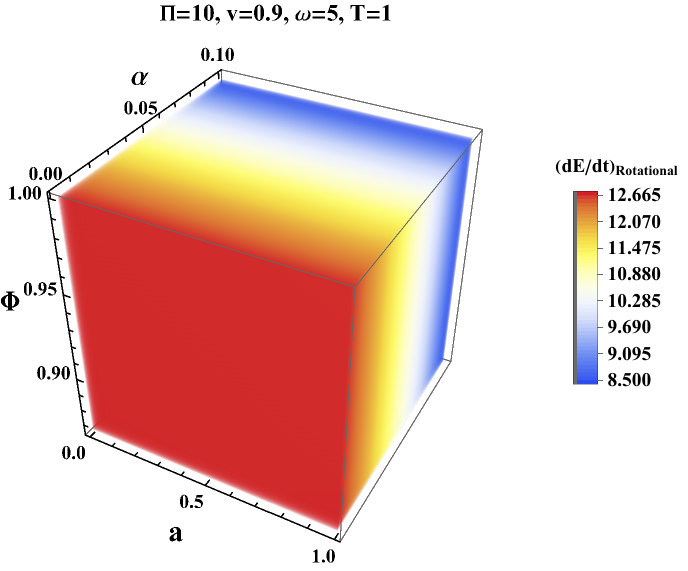}}
	\subfigure[]{\includegraphics[width=0.32\linewidth]{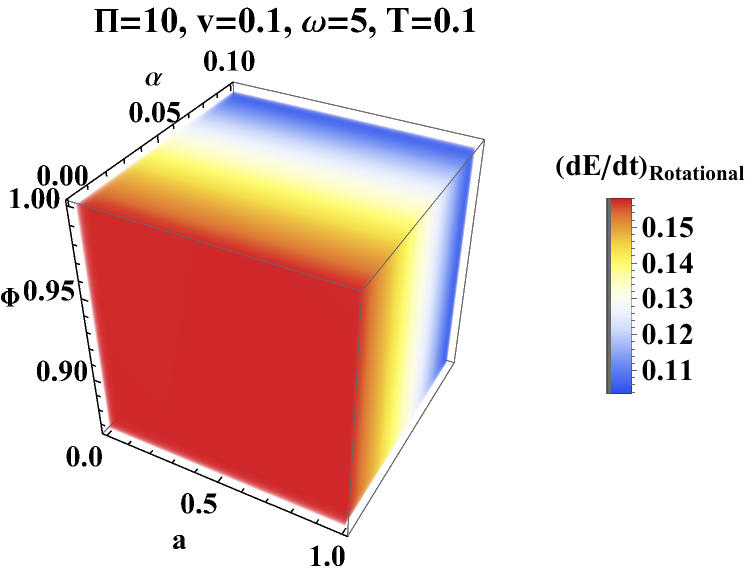}}
	\subfigure[]{\includegraphics[width=0.32\linewidth]{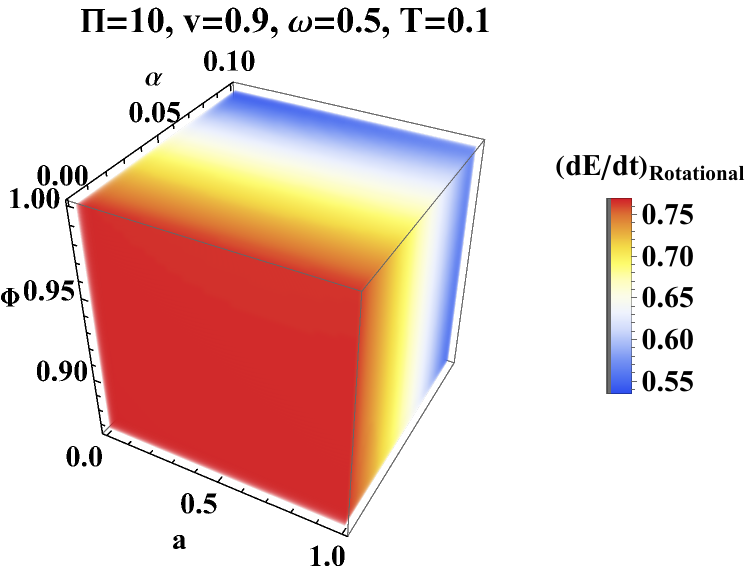}}
	\subfigure[]{\includegraphics[width=0.32\linewidth]{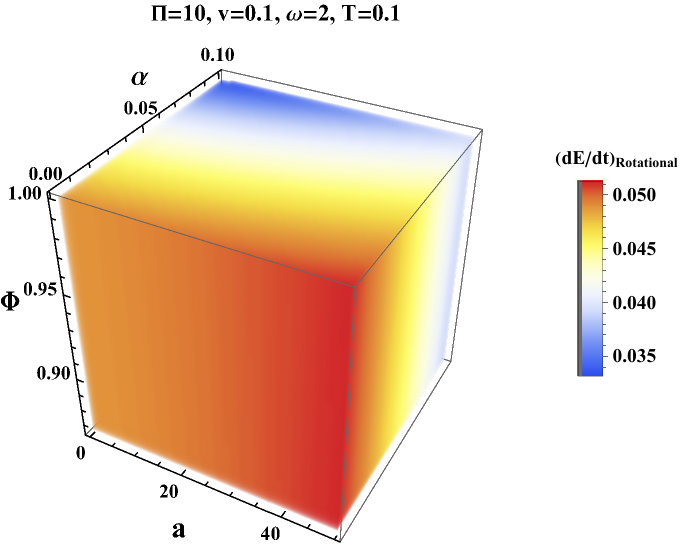}}
	\caption{Density plot representing rotational energy loss ($\frac{dE}{dt}_{Rotational}$) for different set of parameters.}
	\label{fig_energyloss_rot}
\end{figure}

In this section, we study the energy loss of the rotating probe quark through rotational, drag or vacuum energy loss. The interaction of the probe quark with the thermal medium in the presence of the higher derivative GB gravity, potential and string cloud leads to the energy loss. In \cite{Pokhrel_2025,Pokhrel_2025a,Athanasiou_2010,Herzog_2007,Chakrabortty2016a}, the energy loss has been holographically studied, by considering the rotating probe motion in the weakly coupled dual gravity background and the rotational energy loss is given as,
\begin{equation}
	\left.\frac{dE}{dt}\right|_{rotational} = -\frac{\delta S}{\delta (\partial_\sigma X^0)} = \Pi_t^\sigma=\frac{h(u_c)}{2\pi \alpha_t u_c^2},
\end{equation}
where, $u_c$ is the solution of the equation $h(u)-\rho^2\omega^2=0$. Here, $\rho(u)$ is the radius of the probe string at distance $u$ from the boundary and $h(u_c)$ is the value of the metric function at the critical or turning point $u_c$. From the above equation, the energy loss is plotted for range of parameters $a$, $\alpha$, $\Phi$ and different values of temperature, velocity and angular frequency in figure (\ref{fig_energyloss_rot}).
From figure (\ref{fig_energyloss_rot}), it is observed that the rotational energy loss slightly gets reduced with temperature (subfigure a and b) and GB coupling, whereas it gets enhanced with increase in velocity (subfigure a and c), angular frequency (subfigure a and d), flavor density and potential (subfigure e).\\

Further, pure drag energy loss has been studied which is given by,
\begin{equation}
	\left.\frac{dE}{dt}\right|_{drag} = -\frac{\delta S}{\delta (\partial_\sigma X^1)} = \left.\frac{h(u_c)}{2\pi\alpha_t u_c^2}\right|_{drag},
\end{equation} 
where $u_c$ is the solution of the equation \cite{Pokhrel_2025,Chakrabortty2016a},
\begin{equation}
	h(u_c)-v^2=0,
\end{equation}
\begin{figure}[!h]
	\centering
	\subfigure[]{\includegraphics[width=0.32\linewidth]{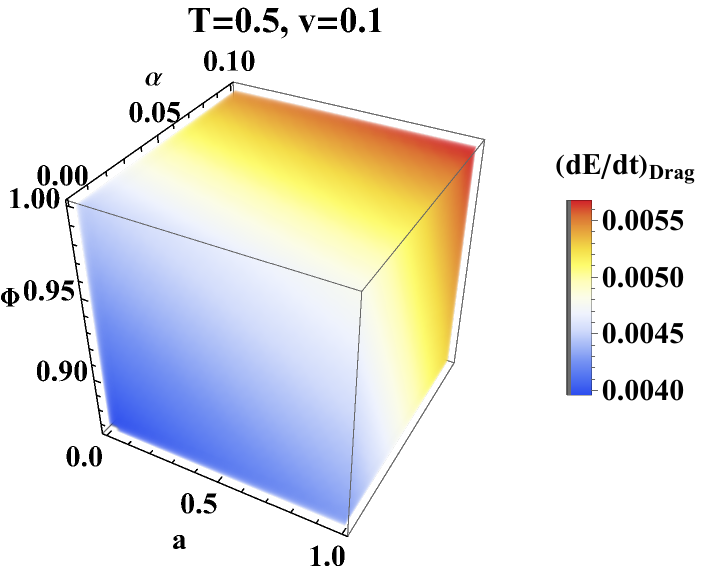}}
	\subfigure[]{\includegraphics[width=0.32\linewidth]{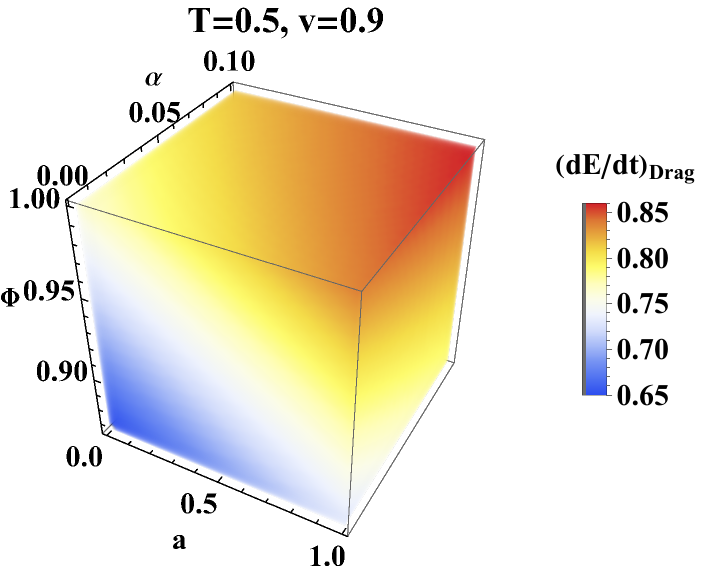}}
	\subfigure[]{\includegraphics[width=0.32\linewidth]{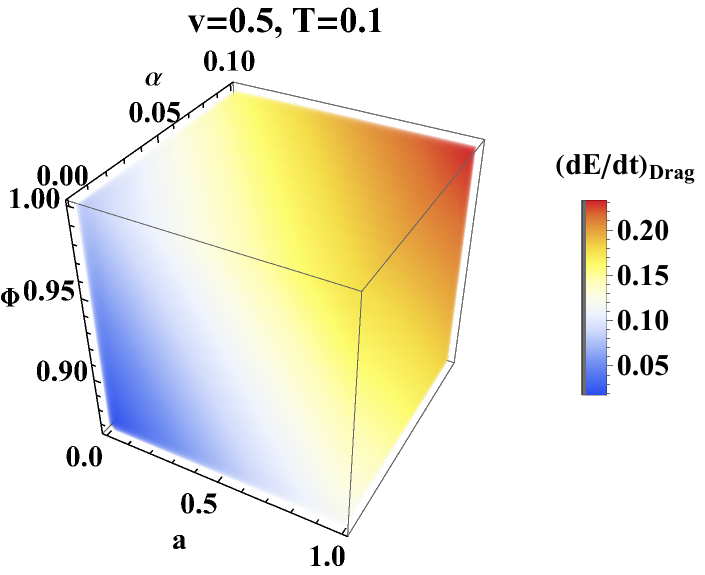}}
	\subfigure[]{\includegraphics[width=0.32\linewidth]{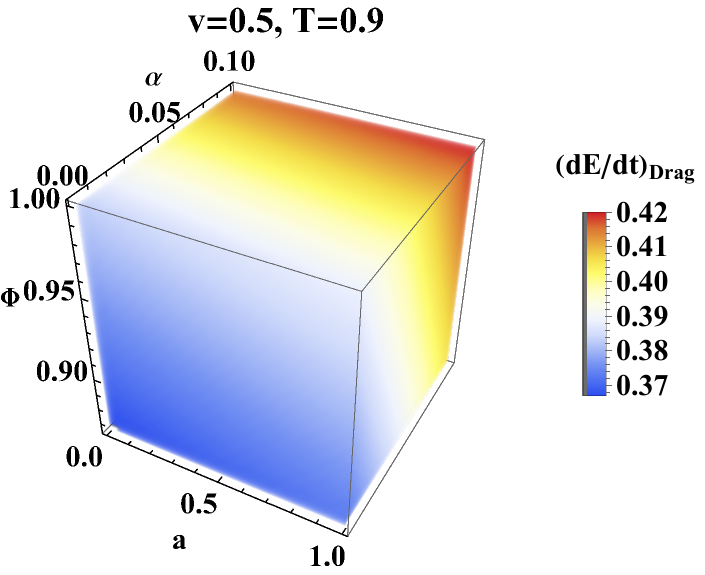}}
	\subfigure[]{\includegraphics[width=0.32\linewidth]{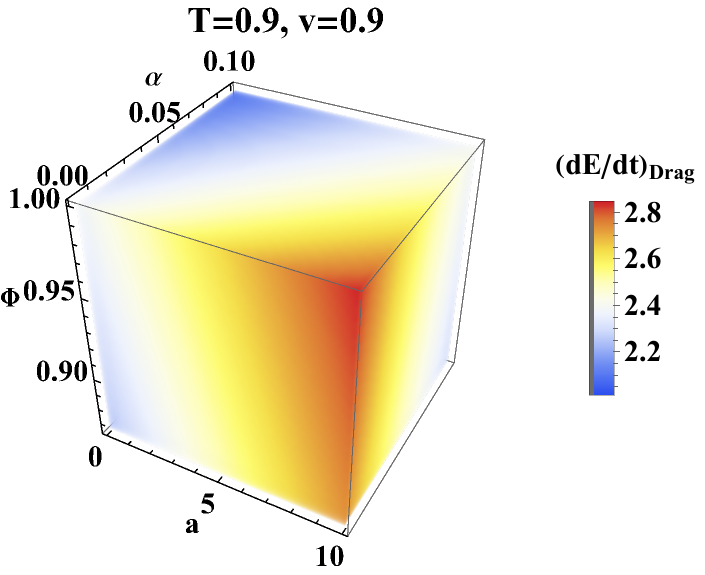}}
	\caption{Density plot representing drag energy loss for different set of parameters.}
	\label{fig_energyloss_pureDrag}
\end{figure}
Similar to rotational energy loss, the pure drag energy loss is plotted for range of parameters in figure (\ref{fig_energyloss_pureDrag}). With increase in velocity (subfigure a and b), temperature (subfigure c and d), flavor density, GB coupling and potential the drag energy loss gets enhanced while for high temperature and velocity increasing GB coupling decreases the drag energy loss (subfigure e).\\

The pure vacuum energy loss has been studied in \cite{Mikhailov2003}, which is given as,
\begin{equation}
	\left.\frac{dE}{dt}\right|_{vacuum} \approx \frac{v^2 \omega^2}{(1-v^2)^2}.
	\label{eq_vacuum_energy_loss}
\end{equation}
It is clearly observed from equation (\ref{eq_vacuum_energy_loss}), that the vacuum energy loss depends only on the velocity and angular frequency of the probe quark and not on the values of the parameters as discussed in \cite{Pokhrel_2025,Pokhrel_2025a} hence we are not repeating the plots here.
\section{Conclusion and Discussion}\label{sec_conclusion}
In this work, we have investigated various transport properties of the baryon rich back-reacted thermal plasma with finite 't Hooft coupling correction dual to charged AdS black hole with higher derivative GB gravity and string cloud. Specially, the transport properties such as drag force, jet quenching parameter of a probe string,  screening length of the $q\bar{q}$ pair in both perpendicular and parallel orientations, radial profile and energy loss of the rotating probe string have been studied.

The drag force experienced by a moving probe quark is enhanced with the increase of temperature, probe velocity, flavor density and baryon potential.  At high temperature and velocity, drag force decreases with GB coupling, otherwise it increases. Whereas, the jet quenching parameter raises with the all parameters.

For both the orientations, the screening length gets reduced with increment in rapidity parameter, temperature, flavor density, baryon potential and finite 't Hooft coupling.  Further, the screening length in parallel orientation is greater than that of perpendicular orientation for the same set of parameters indicating more stable bound state configuration in parallel orientation.

Then, the radial profile of the constantly rotating probe string has been analyzed. It is observed that the radial profile reduces with the increase of baryon potential, flavor density, temperature and angular frequency, whereas it gets enhanced with the increase of conjugate momentum and finite 't Hooft coupling.

Finally, the energy loss of a rotating probe quark has been investigated. An increment in temperature and GB coupling leads to a slight reduction in the rotational energy loss, whereas it enhances for an increment in velocity, angular frequency, flavor density and potential.  A similar qualitative behaviour is observed for the pure drag energy loss, except that, in this case, an increase in both velocity and temperature leads to an enhancement of the drag energy loss. In contrast, the pure vacuum energy loss depends solely on the velocity and angular frequency of the probe quark and is independent of the remaining parameters.

The overall observation regarding the transport properties of the baryon rich back-reacted thermal plasma with finite 't Hooft coupling correction dual to charged AdS black hole with higher derivative GB gravity and string cloud is consistent with the previous findings of \cite{Caceres2006,Gubser_2006,Caceres2006a,Nakano_2007,Chakrabortty2011a,Chakrabortty2016a,Chen:2024epd,Pokhrel_2025,Pokhrel_2025a,Zhu:2024ynp,Zhang_2024,Hou_2021}. Since, chiral symmetry breaking is not considered in the present setup, our results should be interpreted strictly within the domain of this model. The present study may be further extended to other transport properties such as entanglement entropy, photon and dilepton production rates, etc., in the same back-reacted thermal plasma.

\section{Acknowledgement}
RP and IKPC would like to thank the T.M.A. Pai research grant provided by Sikkim Manipal Institute of Technology, Sikkim Manipal University (SMU). KPS would like to thank MoTA, Govt. of India for the research fellowship through NFST. RP would like to thank B. Sharma for providing valuable feedback on the manuscript. The authors are grateful to reviewer for the valuable and detailed comments which led to substantial improvement of the manuscript.

\printbibliography
\end{document}